\documentclass[prd,a4paper]{revtex4}
\begin{document}
%\draft
\title{The general Penrose inequality: lessons
from numerical evidence.}
\author{   Janusz Karkowski$^{+}$ and  Edward Malec$^{+1}$}
\address{ $^+$ Institute of Physics,  Jagiellonian University,
30-059  Cracow, Reymonta 4, Poland.}
\address{$^1$ Erwin Schr\"odinger International Institute, Boltzmanngasse 9, A-1090
Vienna, Austria.}

\begin{abstract}
Formulation of the Penrose inequality becomes ambiguous when the past
and future apparent horizons do cross. We test numerically several natural
possibilities of stating the inequality in punctured and boosted
single- and double- black holes, in a Dain-Friedrich  class of   initial
data and in conformally flat spheroidal data.
The Penrose inequality holds true in vacuum configurations for the  outermost element
amongst the set of disjoint future and past apparent horizons (as expected)
and (unexpectedly) for each of the  outermost past and future apparent
horizons, whenever  these two  bifurcate from an outermost minimal surface,
regardless of whether they  intersect or remain disjoint.
In systems with matter the conjecture breaks down only if matter does not
obey the dominant energy condition.
\\

Dedicated to Andrzej Staruszkiewicz
on the occasion of   his 65$^{\rm th}$ birthday.
\end{abstract}

\maketitle
\section{Introduction.}

The Penrose-Hawking  (\cite{Penrose64}, \cite{Hawking}) singularity theorems point at the
incompleteness of the classical general relativity. The cosmic censorship
hypothesis \cite{Penrose69} can be regarded as an attempt to contain the
damage, by demanding that the {\bf genuine}
singularities are hidden within the black holes.  While it is still not
clear whether the cosmic censorship hypothesis holds true, there is no
doubt that it has shaped the research field and led to many important
results concerning evolving systems (for a recent review see \cite{Chrusciel}).
Penrose invented in 1973 an inequality that can  constitute a necessary
condition for the validity of the cosmic censorship (\cite{Penrose73},
\cite{Gibbons}). The Penrose inequality  has been recently proved,
following a  scenario suggested in 1973  by Geroch \cite{Geroch}
in the important riemannian case (\cite{Huisken},
\cite{Bray}), that can be roughly described as a momentarily static data
set of Einstein evolution equations. It was known for a long time to hold
in spherically symmetric systems \cite{Jang}; a later  analysis   allowed
one to elucidate the problem of the needed energy conditions
and show that it is independent of the
foliation conditions (\cite{ME94}, \cite{Hayward}).
There exist analytical scenarios for the proof
of the general Penrose inequality (\cite{Frauendiener}  and \cite{MMS})
but their validity is not proven, due to the technical complexity.

While numerics cannot per se produce a proof, it may disprove  the
hypothesis or, more likely,  be of help in finding its
correct formulations. The latter  is the main goal of this paper -- Sec. 3
presents  possible wordings of the Penrose inequality
that are checked in later sections in  examples representing
several classes of initial data.
The obtained results can be described as  bringing some surprises.
In the spherically symmetric spacetimes the future and past apparent
horizons are disjoint and  it is well known that the Penrose inequality is
valid only for the outermost of all apparent horizons (later on
abbreviated as AH) \cite{ME94}, assuming
an energy condition. In nonspherical spacetimes   the two AH's
can  intersect in most space-time foliations, including the maximal ones.
Notable exceptions are the polar gauge foliations in which the two horizons
cannot be separated (although they can bifurcate), but  their existence
status is unclear. The apparent ambiguity would be resolved by
accepting the proposal of Horowitz \cite{Horowitz} that one should take
a   surface of a minimal area enclosing all AH's.
It is unexpected   in this context that
whenever the outermost past and future apparent horizons bifurcate
(in a sense specified in Sec. 5) from an outermost minimal surface, then
the inequality holds true for each of them. This remark applies to either
crossing or disjoint AH's,   in all nonspherical vacuum
configurations tested by us. One can convert this phenomenological
observation into a local analytic proof, as sketched in Sec. 5.

      The order of  remainder of this
paper is following. Next section brings the initial constraint equations
and a brief description of the conformal method of constructing initial data.
Several versions of the Penrose inequality are given  in Sec. 3.
Sec. 4 briefly describes the relevant numerical methods.  Obtained results
are reported in Secs 5 -- 7. They support the various versions
of the inequality  for the vacuum initial data in Secs 5
(punctured-Bowen-York data) and 6 (punctured Dain-Friedrich data).
Section 7 deals with non-vacuum spheroidal initial data; in this case the
Penrose  inequality can be broken, if the dominant energy condition is not
valid.   Last section presents main conclusions.

\section{Einstein  constraint equations.}

Let $\Sigma $ be an asymptotically flat
Cauchy hypersurface endowed with  an internal metric $g_{ij}$,
the scalar curvature $R$ and the extrinsic curvature $K_{ij}$. The initial constraint
equations read \cite{MTW}
\begin{eqnarray}
R&=&16\pi \rho +K_{ij}K^{ij}-\left( K_i^i\right)^2
\nonumber \\
&& \nabla_i\left( K^i_j- g^i_jK^l_l\right) = 8\pi j_j,
\label{1.1}
\end{eqnarray}
where $\rho $ and ${\vec j}$ are the mass density and current density of initial material
fields. In the case of maximal slicing condition, $K_i^i=0$, the initial data
can be found by  the conformal method \cite{York1973}. In what follows
we analyze conformally flat  classes of solutions,  corresponding to vacuum
and  spheroidal  systems with matter. The metric reads $g_{ij}=\phi^4  \hat g_{ij}$ where
$\hat g_{ij}$ is the Euclidean metric.

The global energy-momentum can be found from standard formulae
\begin{eqnarray}
E&=&{-1\over 2\pi } \int_{S_{\infty }}d^2S^i\nabla_i\phi ,
\nonumber \\
P_j&=&{1\over 8\pi } \int_{S_{\infty }}d^2S^iK_{ij}.
\label{1.2}
\end{eqnarray}
The asymptotic mass is given by $m=\sqrt{E^2-P_iP^i}$.

An apparent horizon will understood later on
 as a two-dimensional surface $S$ lying in $\Sigma $
with a normal $t$ satisfying one of the two  equations
\begin{equation}
\theta_{^+_-}\equiv \nabla_it^i {_+^-}K_{ij}t^it^j=0,
\label{1.3}
\end{equation}
where the sign $^+$ and $_-$ corresponds to the past and future
apparent horizons, respectively and $\theta $'s are known as optical scalars.

\section{Formulation of the Penrose inequality.}

The Penrose inequality is expected to hold only for the outermost AH.
As explained in \cite{ME94}, in the case of spherical symmetry:
{\it Consider the outermost future trapped surface, the (future) apparent
horizon, call it $S$. Let us assume that $S$ is outside the outermost past
trapped surface. In other words, we assume $\theta_+ (S) = 0$ and that
both $\theta_+ $ and
$\theta_-$ are positive outside $S$}. A simple  analytic argument shows
the validity of the following inequality
\begin{equation}
 m\ge \sqrt{S_H\over 16 \pi },
 \label{1.4a}
\end{equation}
provided that the dominant energy condition is satisfied by matter located
outside $S$. $S_H$ in this formula is the area of $S$. As stressed
in the quoted paper, {\it Of course, an identical argument works if
the outermost trapped surface is a past apparent horizon.}

In  the spherically symmetric geometries optical
scalars have the same level sets, since both of them are constant
on centered spheres. Thus the spheres surrounding the outermost
AH have positive $\theta_- $ and $\theta_+$ and i) possess larger area
than the AH. Obviously ii) the future and past horizons do not cross.
Therefore there  is no ambiguity in defining the Penrose inequality
and -- since the cosmic censorship hypothesis asserts that AH's are enclosed
by the event horizon that asymptotically evolves to
a  Schwarzschild or Reissner-Nordstroem black hole horizon
-- it can  be regarded as the necessary condition for the cosmic censorship
\cite{Penrose73}.

None of the features i) and ii) becomes obvious in the general nonspherical
case, even if  our liberal definition of AH's is replaced
by the more stringent one  due to Penrose. Let us recall that in
\cite{Penrose64} future trapped surface are assumed to
have -- in our terminology  -- a positive  scalar $\theta_- $
and a negative scalar  $\theta_+$. That is,  each of the two
beams of null geodesics emanating orthogonally
outward and inward from a trapped surface, is convergent. Consequently,
a future apparent horizon (understood as the outermost boundary in the set
of all future trapped surfaces) has vanishing $\theta_+$ but non-negative
$\theta_-$. The analogous situation (but with optical scalars reversing
their roles) takes place for past apparent horizons.
Thus the  picture of AH's that emerges here
resembles that of spherically symmetric geometries.
There exist level sets of, say $\theta_+<0$, such that $\theta_- >0$ (and
conversely). The two optical scalars do not possess common level sets, but
the sign of one of them is  controlled  on the level set of the other.
This can happen only if one   matches in a suitable way the
choice of both a Cauchy hypersurface
and  of the two-dimensional foliation within this slice. (This is
inherent also to  the scheme of the proof of the Penrose inequality
that is proposed in \cite{MMS}). One can
find a two-surface $S$ to be such  an AH in one particular foliation,
but that may not be true in other space-like slices.
Even with this stringent definition,   the future and past AH's
can intersect and there may exist surfaces of a smaller
area surrounding them as pointed out by Horowitz \cite{Horowitz}.

In the rest of this paper by AH's are understood two-surfaces satisfying
one of the conditions of (\ref{1.3}), which might   be weaker
than the notion employed in the singularity theorems (but see a discussion
following the point ii) below).
 That means  that the failure of a particular version (or all of them)
formulated below of the Penrose inequality does not necessarily
negate the  the cosmic censorship hypothesis (CCH).
And conversely, their validity lends even more credence in
 CCH. It is not without significance that such AH's are easier to find
numerically than the    standard objects defined by Penrose.

The three versions of the Penrose inequality read as follows,
 (assuming the dominant energy condition \cite{Hawking} for nonvacuum initial data)

i)
The minimalistic   one (PIM henceforth); it was borrowed from a proposition
first put forward by Horowitz \cite{Horowitz}.
 {\it The surface $A_M$ of a smallest area $S_M$ surrounding
regions with   horizons}  satisfies the inequality
\begin{equation}
 m\ge \sqrt{S_M\over 16 \pi }.
 \label{1.4}
\end{equation}
It appears  in many of the numerical cases reported later that $A_M$ coincided
with that constructed from apparent horizons (see ii)), but in a number of
configurations it consisted also of segments of minimal surfaces.
The existence of configurations having  portions of minimal
surfaces extending outside AH's, means that  it is not  excluded that
the actual area of an event horizon -- if there is one --
is   smaller than that of the AH. In such a case  PIM  constitutes the
  necessary condition for the validity of CCH.

ii) The   standard  one  (PIS later on).  {\it The closed 2-surface $A_H$  is either the
outermost apparent horizon (if AH's do not
intersect) or a   union of segments of outermost future
and/or past apparent horizons.}
  Then its area $S_H$  satisfies the inequality
\begin{equation}
 m\ge \sqrt{S_H\over 16 \pi }.
 \label{1.5}
\end{equation}
$A_H$ does not manifestly satisfy the assumptions of the singularity theorems,
but its importance lies in the fact that it may  do so
in another  foliation (say, polar gauge one). The heuristic argument is as
follows. The product of two optical scalars is i) invariant and ii)
 vanishes on $A_H$.
If there exists  a local boost to a  polar gauge foliation (that is,
a foliation with $\theta_-=\theta_+$; apparent horizons correspond  here to minimal
surfaces) of a space-time, then on a polar gauge slice the  two-surface
$A_H$ would become an  apparent horizon (that is, the outermost minimal
surface) in the sense of Penrose, and then the CCH
demands the existence of an event horizon. The area of the
intersection of the event horizon with the actual polar gauge slice would
have to be bigger than of $A_H$. Accepting that, PIS seems to be just right
one, as a necessary condition, from the point of view of CCH.
Unfortunately, there is a gap in the argument. Namely, there is no
possibility to rule out the existence of minimal surfaces that extend
outward of the outermost apparent horizon. There are reasons to expect
(basing on the analogy to spherical symmetry) that the polar gauge slice
does  not penetrate regions with  minimal surfaces and the surface $A_H$ would not
be seen on the slice.

iii) In the cases with intersecting apparent horizons $A_A$'s we compared
also their area related quantities $\sqrt{S_A\over 16 \pi }$ with the
asymptotic mass. Invariably it was found that
\begin{equation}
 m \ge \sqrt{S_A\over 16 \pi }
 \label{1.5a}
\end{equation}
for each of the horizons, and with a significant safety margin.
There is no obvious reason why the area $S_A$  should be bigger than
$S_H$, but this is what was found to be true in all analyzed examples.
That suggests that iii) is stronger than PIS  and it comes  as a surprise
that  numerics supports the inequality (\ref{1.5a}).

Data concerning the  two stronger (ii) and iii)) of the
above conjectures are given in sections 5 and 6; their validity implies PIM.
Only in the first part  of Sec. 5 we report data concerning
the version PIM, to show that the results   are close for  all three statements.
The horizons do not cross in the case of spheroidal initial data
and there is no minimal surface outside the outermost AH; therefore
conjectures PIS and PIM do coincide in the   examples considered in Sec. 7.

\section{Description of numerical methods.}

In the conformal method and for vacuum conformally flat initial data, one first solves
analytically  the equation $\hat \nabla_i\hat K_j^i=0$ (the covariant derivatives
are in the Euclidean metric) and then the Lichnerowicz-York equation
\begin{equation}
\Delta \phi = -{ \hat K_{ij}\hat K^{ij}\over 8} \phi^{-7},
\label{0.1}
\end{equation}
with the flat laplacian $\Delta $. This is an example of a weakly nonlinear elliptic equation;
its leading derivatives are linear (hence the equation is quasilinear) and the nonlinearity is
rather weak (c.f. negative powers of  the conformal factor $\phi $). Due to the cylindrical symmetry,
we search for $\phi $ as a function of the angle $\theta $ and the coordinate radius $r$.
It is necessary to map the problem into one with finite domain; thus   the radius $r$ is replaced
by another independent variable $v= r/(1+r)$. Adopting $x=\cos \theta $, one has to solve
(\ref{0.1}) in the rectangular $-1\le x\le 1, 0\le v\le 1$. It is solved iteratively by the
standard Newton method on the lattice up to $200\times 5000$ points. Due to the weak nonlinearity of
the Lichnerowicz-York equation, it was enough to apply at most 4-5 iterations. We used four different solvers,
in particular  the MUMPS  \cite{MUMPS} and HYPRE \cite{HYPRE} ones.

The apparent horizon equation (\ref{1.3}) becomes in our context a nonlinear ordinary equation,
for the function $r(\theta )$. This is a classical two-point problem (see a discussion in \cite{kark92} in a similar
context) with $dr/d\theta |_{\theta =0}=dr/d\theta |_{\theta =\pi }=0$. It is solved by
the standard shooting method. We resorted to two numerical packages, ODEPACK \cite{ODEPACK} and SUBPLEX
\cite{SUBPLEX}. The needed extrapolation of the formerly found solution $\phi $ has been done with the
help of the bilinear interpolation \cite{Teukolsky}. The bilinear method appeared  entirely satisfactory, due to the high density
of our numerical lattice.

\section{Boosted punctured data.}

Assume   $\hat P$ to be a constant vector
and $\hat n$ - a unit normal to a metric sphere in the Euclidean geometry. Assume standard
spherical coordinates $r, \theta $ and $\phi $. One can easily check
that the   extrinsic curvature
\begin{equation}
K_{ij}={3\over 2r^2 \phi^2}\left( \hat P_i\hat n_j + \hat P_j\hat n_i -\left( \hat g_{ij}
-\hat n_i\hat n_j\right) \hat P_l\hat n^l \right)
\label{2.1}
\end{equation}
satisfies the momentum part of Eq. (\ref{1.1}) with the vanishing current (that is the boosting part of the
Bowen York initial extrinsic curvature \cite{Bowen}). The hamiltonian
constraint (the first equation in (\ref{1.1})) reads now, assuming vacuum case  ($\rho =0$)
and aligning the $z$-axis along $\hat P$,
\begin{equation}
\Delta \phi = -{9  (\hat P)^2\over 16r^4}\left( 1+2 \cos^2\theta \right) \phi^{-7},
\label{2.2}
\end{equation}
where $\Delta $ is the flat laplacian.  For these boosted   data  one
obtains the global momentum $P_i=\hat P_i$, the global energy $E$ is given by (\ref{1.2})
and the asymptotic  mass reads $\sqrt{E^2-P^2}$.

There are two established ways of solving the resulting (Lichnerowicz-York)
equation. i) In the first approach, that takes care about  the global topology
of the manifold (the so-called conformal imaging method \cite{Bowen}; that
actually requires the use of a larger  set of extrinsic curvature data),
Bowen and York solve Eq. (\ref{2.2}) outside $r\ge a$, assuming that
the sphere $r=a$ is a minimal surface and that at infinity the conformal
factor $\phi $ goes to 1.
ii) In the second approach, the puncture method, one splits  $\phi $
into two parts, $\phi = 1+ m_1/(2r )+\phi_1$  and finds a solution
$\phi_1$ in the whole Euclidean space, demanding that
at infinity $\phi_1\approx d_1/(2r)$ \cite{Brugmann}; here $m_1$ and $d_1$
are some constants, that are related to the global energy of the manifold.

In this paper we use the second approach in order to treat  vacuum
configurations with one or two black holes.
The data for three exemplary configurations (chosen from a much bigger sample)
with a single black holes are presented
in the first table. The first column is the parameter $m_1$ appearing in the
preceding  splitting, the second column is the linear momentum  $\hat P$,
and the third, fourth and fifth columns, respectively are the global mass $m$
and the "horizon" masses $m_M =\sqrt{S_M\over 16 \pi }$,  $m_H =
\sqrt{S_H\over 16 \pi }$ and $m_A = \sqrt{S_A\over 16 \pi }$.
In the case of single boosted black holes
the asymmetry causes the horizons to intersect the minimal surface, and
this is why we consider   the case PIM with $A_M$.
The surfaces in the fourth and fifth columns
do not coincide with  apparent horizons, but   consist  of two ($A_H$)
parts (of an apparent horizon to the future and to the past) or  three
segments ($A_M$) (of an apparent horizon to the future, a minimal surface
and an apparent horizon to the past) -- as explained in Sect. III.
 It happens  that the areas $S_A$  of the  apparent horizons to the past and
  to the future are equal; the sixth column brings
corresponding data which (unexpectedly) obey the Penrose inequality.
Each row describes a  different configuration.
%
%\begin{equation}
\begin{center}
\begin{tabular}{|c|c|c|c|c|c|}
   \hline
   % after \\: \hline or \cline{col1-col2} \cline{col3-col4} ...
   $m_1$ & $\hat P $ & m &$m_M$ & $m_H$ & $m_A$ \\
   \hline
   8 & 2 & 8.061855 & 8.059402 &8.059426 & 8.05948097\\
  \hline
   4 & 2 & 4.122407 & 4.110666 &4.110751 & 4.11093771\\
 \hline
   4 & 5 & 4.707092 & 4.499335 &4.500002 & 4.50143757\\
\end{tabular}
\end{center}
%\end{equation}
%
In this case areas  of the past and future AH's are equal.
It is clearly  seen that all versions, the weaker  (PIM) and   the stronger
(PIS) as well the last one (iii) of the  Penrose inequality are satisfied.
It is noticeable that the   areas  of $A_M$, $A_H$ and $A_A$ are very close.
The  fact  of interest is the numerical evidence for the existence
of parts of minimal surfaces that extend outward of outmost apparent horizons.

The corresponding results for two black hole configurations are comprised
in the next table. Now  the puncture method  requires that
$\phi = 1 + \sum_{i=1}^2m_i/(2|{\vec r}-{\vec r_i}|)+\phi_1$
(the two  black holes are located at ${\bf r_i}$, i=1,2). The extrinsic
curvature reads
\begin{equation}
K_{ij}={3\over 2r^2 \phi^2}\sum_{s=1,2} \left( \hat P_i^{(s)}\hat n_j^{(s)} + \hat P_j^{(s)}
\hat n_i^{(s)} -\left( \hat g_{ij}
-\hat n_i^{(s)}\hat n_j^{(s)}\right) \hat P_l^{(s)}\hat n^{(s)l} \right) ,
\label{2.3}
\end{equation}
where $\hat P^{(s)}$ aligned along the $z$-axis
 can  be interpreted as the linear momentum of the
$s-th$ black hole and ${\vec n}^{(s)}=({\vec r}- {\vec R}_s)/|
{\vec r- \vec R_s}|$. As before, one finds a solution
$\phi_1$ of the Lichnerowicz-York equation in the whole Euclidean space,
demanding that
at infinity $\phi_1\approx d_1/(2r)$ \cite{Brugmann}; as before  $m_1$ and
$d_1$ are some constants, that are related to the global energy of
the manifold. The first   and second columns describe parameters
("mass" $m_1$ and "momentum" $\hat P^{(1)}$
of the first black hole, the third and fourth columns give the same
information about the second black hole. The fifth and sixth columns,
respectively are the global mass and the areal mass $m_H=\sqrt{S_H\over
16 \pi }$. $S_H$ is the area of $A_H$, the 2-surface constructed according
to the recipe ii) of Sec. II.
The 2-surface $A_H$ surrounds   both black holes (located at $r=1$ and
$\theta =0$ or $\theta =\pi $). In two cases (fifth and ninth) the past and
future AH's do cross; the seventh column presents relevant values of
$m_A=\sqrt{S_A\over 16 \pi }$, where $S_A$ is the larger of the two areas
in question. They satisfy all formulations of
the inequality.
\begin{center}
\begin{tabular}{|c|c|c|c|c|c|c|}
   \hline
   % after \\: \hline or \cline{col1-col2} \cline{col3-col4} ...
   $m_1$ & $\hat P^{(1)} $ & $m_2$ & $\hat P^{(2)} $ &m & $m_H$ & $m_A$\\
   \hline
   5 & -5 & 5& 5    & 10.040901 & 10.033511 & \\
   \hline
   5 & -10 & 5& 10  & 10.159675 & 10.132316 & \\
   \hline
   5 & -10 & 5& 8   & 10.176694 & 10.153778 & \\
   \hline
   5 & -10 &5 & 5   & 10.379353 & 10.347249 & \\
   \hline
   5 & 5 & 5  &5    & 11.146633 & 10.916395  & 10.916399\\
   \hline
   5 &-0.25&5 &0.25 &10.000103 &9.999632 & \\
   \hline
   5 & -2.5&5 & 2.5 &10.010290 &10.008087 & \\
   \hline
   4 & -1 & 5& 1.5  & 9.006743 &9.005594 & \\
   \hline
   5 & -2.5&5 & 0.5 & 10.052687 &10.051593 & 10.051593 \\
\end{tabular}
\end{center}
In the remaining seven cases the past and future horizons do not cross
and there is a minimal surface in between them. Surprisingly -- and in a sharp contrast
with the corresponding case in spherically symmetric configurations -- the
Penrose inequality   is valid simultaneously for past and future AH's.
Another interesting observation is that, in all   cases,
  when the parameter $\hat P$
tends to zero then AH's tend to the minimal surface. In this sense, the
AH's bifurcate from the minimal surface.  This
is true, as matter of fact, in all numerical examples studied in Secs. 5 and 6.
It happens, that there always exists at least one minimal surface;
those AH's horizons that bifurcate from the outermost one do satisfy
the inequality. On the other hand, AH's   that branch from
an innermost minimal surfaces (there are several such cases  in
our sample of data)  do break all aforementioned  versions.

We show below  an analytic argument that these   observations
remain true (with some reservations)
for  initial data with AH's that arise from small perturbations
of data with minimal surfaces.

{\bf Theorem.} Let $l$ be a real parameter, $l\hat K_{ij}$
 be the Bowen-York or Dain-Friedrich extrinsic curvature  (multi-puncturized) and
 $\phi_l $  be a solution of the Lichnerowicz-York  equation on $R^3$
\begin{equation}
\Delta \phi_l = -{ l^2\hat K_{ij}\hat K^{ij}\over 8} \phi^{-7}_l,
\label{A.1}
\end{equation}
Then $\phi_l$, $K_{ij}=\hat K_{ij} /\phi^2_l$ constitute   initial data
of the Eintein equations; assume that for each  $l$ there exist apparent horizons
that in the limit $l\rightarrow 0$ coincide with a nonspherical outermost minimal surface
$S_0$. Then there exists  $l_0$ such that for $|l|<l_0$ the Penrose inequality
is satisfied.

{\bf Sketch of the proof.} In the case of a nonspherical surface $S_0$
given by $r=r_0(\theta )$, one has a strict inequality,
$\epsilon \equiv m_0-\sqrt{A_0/(16\pi)}>0$. One can easily show that: i)
(using arguments  of \cite{York1973}) $\phi_l\ge \phi_0$ and   $m_l=m_0+c_1l^2$;
ii) (using the Green function of the flat laplacian) $\phi_l<\phi_0 +c_2l^2$.
Here and below $c_i$ ($i=1, 2,...$)  are some positive constants.
The apparent horizon  equations $\theta_+=0$ or $\theta_-=0$ depend on
$l$ through the extrinsic curvature terms and
the conformal factor $\phi $. Since by assumption horizons bifurcate from $S_0$,
they must be located within annulus $(r_0(\theta, \phi )-c_3l,
r_0(\theta, \phi )+c_4l)$. (This is due to the implicit function
theorem).  At an apparent horizon
one should compare $m_l=m_0+c_1l^2$ with the area of AH's, which is bounded
from above by $A_0+c_5l$. It is clear that by choosing $l_0$ small enough
one can ensure that $m_l\ge \sqrt{A_H\over 16\pi }$.

This proof is insensitive on the sign of $l$ and therefore
it is valid for both past and future apparent horizons. Any attempt to convert
this local result  into global one would have to be preceded by a
careful estimate of the dependence of the location $r(\theta )$ of an AH on
the bifurcation parameter $l$. Notice that
this  theorem does not apply to single-puncture initial data, since in this case
the geometry corresponding to $l=0$ is spherically symmetric and
$\epsilon =0$. On the other hand,  this result should
hold for the two-puncture solutions.
There is also a possibility of generalizing the above onto case with
initial data given in the exterior of a two-surface $S_1$, instead of
$R^3$.

\section{Dain -- Friedrich   conformally flat initial data.}

The main feature of these initial data  is   that the spatial part
of the metric is conformally flat (as before), the momentum flow density
vanishes (again, as before) and the extrinsic curvature is given, in
spherical coordinates, by (the forthcoming formulae are translated from
the language of the Newman-Penrose formalism, originally used in \cite{Dain}).
\begin{eqnarray}
K_i^3&=&0,~~~ for ~i=1,2\nonumber
\\
&&K_1^1={1\over r^3\phi^6}\partial^2_xW \nonumber
\\
&& K_1^2={1\over \sin \theta r^3\phi^6}\partial_r\partial_xW\nonumber
\\
&&K_2^2={1\over r^2\sin^2\theta \phi^6}\left[ \partial_r\left( r\partial_rW\right) +{1\over r} \left( x\partial_xW
-W\right) \right] .
\label{4.1}
\end{eqnarray}
Here $x=\cos \theta $ and $W $ is an arbitrary function of $r$ and $\theta $.  The $K_3^3$ component
can be found from the maximal slicing condition $K_i^i=0$. Let us point out
that this solution generalizes the Bowen-York solution of momentum
constraint; the latter corresponds to a   particular choice of $W $\cite{Karkowski04}.
The data of (\ref{2.1}), for instance, correspond to $W=r\hat P\left(
x^3-3x\right) /2$. Extrinsic curvature  (\ref{4.1}) constitutes a partial
case of the general solution found by Dain and Friedrich in 2001 \cite{Dain}.

The conformal factor $\phi $ satisfies
the Lichnerowicz-York equation $\Delta \phi = -{1\over 8} K_{ij}K^{ij}\phi^5$.
We seek a solution $\phi $, using the puncture method,
of the form $\phi =1 +m_1/(2r) +\phi_1$.

 The    numerical calculations have been  performed in the following cases:

i) $\partial_xW=Pr \left( -x+x^3 \right)$. It is noticeable that here
the global momentum is nonzero. Numerical results are given in the
forthcoming Table.
\begin{center}
\begin{tabular}{|c|c|c|c|c|}
   \hline
   % after \\: \hline or \cline{col1-col2} \cline{col3-col4} ...
   $m_1$ & $ P $ & $m$  & $m_H$& $m_A$\\
   \hline
   4 & 5 &5.622456   &4.296457 &4.296680 \\
   \hline
   4 & 1 & 4.078569  &4.015156 &4.015260 \\
\end{tabular}
\end{center}

ii) $\partial_xW=Pr(1-x^2)(3x^2-1)/2 $. In this case the global momentum vanishes. The Table
presents the obtained results.
\begin{center}
\begin{tabular}{|c|c|c|c|c|}
   \hline
   % after \\: \hline or \cline{col1-col2} \cline{col3-col4} ...
   $m_1$ & $ P $ & $m$  & $m_H$ & $m_A$ \\
   \hline
   4 & 5 &6.073919  &4.366950 & 4.372253 \\
   \hline
   4 & 1 & 4.109128  &4.021109 &4.021207 \\
\end{tabular}
\end{center}
In both cases i) and ii) the two stronger versions,  PIS and iii),
of the Penrose inequality holds true. The 2-surface $A_H$, whose areal mass is depictured
in the last but one column, is built from many sections of the intersecting
past and future horizons;  the number of the intersections seems to depend
(for a given nonzero $P$) on the shape of $W$
as a function of $\theta $. There are more crossings in case i) (three ) than in
the case ii) (only two).

\section{Spheroidal systems with matter.}

Assume a foliation of the Euclidean space by oblate spheroids,
\begin{equation}
{x^2+y^2\over a^2 \left( 1+\sigma ^2 \right) }+{z^2\over a^2 \sigma^2}=1.
\label{5.1}
\end{equation}
The variable  $\sigma $ changes from 0 to $\infty $, and angle variables are
 $\tau $  (changing from  -1 to 1) and    $\phi $ (varies as usual from $ 0 $ to $2\pi $).
Assume that there exists a normal flow of matter with the only nonzero component
\begin{equation}
j_{\sigma }= {1\over 8\pi }\phi^{-6} {\sigma \left( \tau^2-1\right) \over \left( \sigma^2+1\right)
 \left( \sigma^2+\tau^2\right)^{5/2}}.
\label{5.2}
\end{equation}
Let $\hat n_i$ denote the unit normal (in the Euclidean metric) to a spheroid.
The related traceless extrinsic curvature reads
\begin{equation}
K_i^j=-\phi^{-6} \left( \hat n_i\hat n^j -{1\over 3}g^j_i\right)
{1\over \left( \sigma^2+\tau^2 \right)^{3/2}};
\label{1.6}
\end{equation}
this pair, $K_{ij}$ and $j_i$, solves the momentum constraint part of Eq.
(\ref{1.1}). Let us remark, that one could always add to the extrinsic
curvature the  diagonal components $K_r^r=C/(\phi^6r^3),
K_{\theta }^{\theta }=K_{\phi }^{\phi }=-K_r^r/2$ (where $C$ is a constant
and $r=\sqrt{x^2+y^2+z^2}$)  without changing the momentum flow.
We do not do  this, because our primary
intention is to study the influence of the   energy conditions  onto the
validity of the Penrose inequality, and the aforementioned  part of
the extrinsic curvature is irrelevant from this point of view.

The energy density $ \rho $  can be chosen in an arbitrary way,
but the simplest possibility - that eases the analysis of the
energy conditions -  is to assume
\begin{equation}
\rho= C\times \phi^{-8}\times {1\over 8\pi }{\sigma \left( 1-\tau^2\right)
\over \left( \sigma^2+1\right)
 \left( \sigma^2+\tau^2\right)^{5/2}}.
\label{5.3}
\end{equation}
Later we shall put either  $C=1$ - which ensures the dominant energy condition
- or $C=0$, which breaks the energy conditions.
Notice that
\begin{equation}
K_{ij}K^{ij}={2\over 3}\times \phi^{-12} \times {1\over \left( \sigma^2+\tau^2\right)^3}.
\label{5.4}
\end{equation}
The  Lichnerowicz-York equation takes now the form

\begin{equation}
\Delta \phi  = -  {1\over 12} {1\over \left( \sigma^2+\tau^2\right)^3 } \phi^{-7} -
{C\over 4}  {\sigma \left( 1-\tau^2\right) \over \left( \sigma^2+1\right)
\left( \sigma^2+\tau^2\right)^{5/2}}\phi^{-3}.
\label{5.5}
\end{equation}
This equation has been solved    with $\phi $ tending
to 1 at infinity and bearing a constant value at the inner boundary, that is assumed to be
a unit sphere $r=1$  in the background Euclidean geometry. In the examples
shown below, horizons do not intersect and in all cases
the outermost apparent horizon (future or past)   has been located outside the minimal
surface. (There appear also  initial data with intersecting horizons, but they
are not of particular interest.) Its  area $S_H$ enters the forthcoming data through the formula
$m_H=\sqrt{S_H\over 16 \pi }$.

We shall present data corresponding to:

i) $C=1$,
\begin{center}
\begin{tabular}{|c|c|c|}
   \hline
   % after \\: \hline or \cline{col1-col2} \cline{col3-col4} ...
   $\phi (r=1)$ &    $ m$ & $m_H$\\
   \hline
   2.4 &  2.801054 & 2.800734 \\
   \hline
   2.5 &  3.0009593 & 3.0006660 \\
   \hline
   3.5&  5.0004402 & 5,0003797834.\\
\end{tabular}
\end{center}
It is clear that $m_H>m$ and that the version PIS holds true.

ii) $C=0$.   The energy density vanishes and therefore the dominant energy condition is broken.
One expects that the Penrose inequality (PIS) may be broken now, and in fact this is
what happens, albeit only in the first two examples.
\begin{center}
\begin{tabular}{|c|c|c|}
   \hline
   % after \\: \hline or \cline{col1-col2} \cline{col3-col4} ...
   $\phi (r=1)$ &     $m$ & $m_H$\\
   \hline
   2.4 &  2.80004295 & 2.80011252  \\
   \hline
   2.5 &  3.00003632 & 3.000604868  \\
   \hline
   3.5&  5.000009651 & 5.000005477 \\
\end{tabular}
\end{center}
\section{Concluding remarks.}

The weakest form of the  Penrose inequality due to Horowitz is that
$m\ge \sqrt{S_M/16\pi }$, where $S_M $ is the smallest area of a
two-surface encompassing a region with apparent horizons satisfying the assumptions
of the singularity theorems. This paper deals with
three other  formulations, of  which even the weakest  (PIM) is stronger
than the Horowitz's one, because our notion of the outermost apparent
horizon is weaker than that required by the singularity theorems.
Despite this fact,  all investigated statements of the Penrose inequality
are confirmed by our numerical analysis for vacuum initial data
and for  those systems with matter that satisfy an energy condition.
The only negative examples correspond  to data with matter that does not
satisfy an energy condition.

There exist  minimal surfaces in all investigated examples with vacuum; these are
mostly singlets but in a number of cases also doublets. It is observed that
the two (past and future)  AH's, that bifurcate (with the
momentum being the bifurcation  parameter --  see the end of Sec. 5)
from the outermost minimal surface, do satisfy
 -- regardless of whether they cross or do not cross --  all versions of the
Penrose inequality. An analytic proof that this is   true (with some
reservations), at least for initial data with small  extrinsic curvature,
 is sketched at the end of Sec. 5.

In summary, results of this work show that the Penrose inequality does not hold
only when expected not to hold; that
points strongly in favour of the validity of the Penrose's conjecture
in physically interesting cases.

\begin{acknowledgements}
We dedicate this paper to Andrzej Staruszkiewicz on the occasion of his 65th birthday.
This work bases on a talk given by EM at the ESI Workshop on the Penrose Inequality held in 2004
and  it was partly supported by  the KBN grant 2 PO3B 00623.
EM wishes to thank Szymon Leski for  his help in dealing with tables, and Sergio Dain and Marc Mars for a very
useful discussion.
\end{acknowledgements}

\end{document}